\documentclass[useAMS,usenatbib]{mn2e}
\usepackage{url}
\usepackage{graphicx}
\usepackage{verbatim}
\usepackage[usenames]{color}
\usepackage[flushleft]{threeparttable}
\DeclareGraphicsExtensions{.pdf,.png,.jpg,.mps,.eps,.ps}
\usepackage{amsmath}
\usepackage{natbib}
\usepackage{color}
\usepackage{graphicx}
\usepackage{lscape}
\usepackage{gensymb}

\newcommand\nicer{{\it NICER}}
\newcommand\nustar{{\it NuSTAR}}
\newcommand\swift{{\it SWIFT}}

\newcommand\rxte{{\it RXTE}}

\newcommand\inte{{\it INTEGRAL}}

\newcommand\kev{{\rm~keV}}

\newcommand\kms{\ifmmode {\rm~km\ s}^{-1} \else ~km s$^{-1}$\fi}
\newcommand\Hunit{\ifmmode {\rm~km\ s}^{-1}\ {\rm Mpc}^{-1}
        \else ~km s$^{-1}$ Mpc$^{-1}$\fi}
\newcommand\ctssec{\ifmmode {\rm~count\ s}^{-1} \else ~count s$^{-1}$\fi}
\newcommand\ergsec{\ifmmode {\rm~erg\ s}^{-1} \else
        ~erg s$^{-1}$\fi}
\newcommand\funit{\ifmmode {\rm~erg\ s}^{-1}\;{\rm cm}^{-2} \else
        ~ergs s$^{-1}$ cm$^{-2}$\fi}
\newcommand\phflux{\ifmmode {\rm~photon\ s}^{-1}\;{\rm cm}^{-2}
        \else   ~photon s$^{-1}$ cm$^{-2}$\fi}
\newcommand\efluxA{\ifmmode {\rm~erg\ s}^{-1}\;{\rm cm}^{-2}\;{\rm
        \AA}^{-1} \else ~erg s$^{-1}$ cm$^{-2}$ \AA$^{-1}$\fi}
\newcommand\efluxHz{\ifmmode {\rm~erg\ s}^{-1}\;{\rm cm}^{-2}\;{\rm
        Hz}^{-1} \else ~erg s$^{-1}$ cm$^{-2}$ Hz$^{-1}$\fi}
\newcommand\cc{\ifmmode {\rm~cm}^{-3} \else cm$^{-3}$\fi}
\newcommand\FWHM{\ifmmode {\rm~FWHM} \else ${\rm~FWHM}$\fi}
\newcommand\Msun{\ifmmode M_{\odot} \else $M_{\odot}$\fi}
\newcommand\Lsun{\ifmmode L_{\odot} \else $L_{\odot}$\fi}

\newcommand\hbeta{\ifmmode {\rm H}\beta \else H$\beta$\fi}
\newcommand\Kalpha{\ifmmode {\rm K}\alpha \else K$\alpha$\fi}
\newcommand\nh{\ifmmode N_{\rm H} \else N$_{\rm H}$\fi}

\usepackage{graphicx}
\usepackage{url}
\usepackage{verbatim}
\usepackage{color}

\title[\nustar{} and \nicer{} view of of XTE J1739-285]{Evidence of hard power-law spectral cutoff and disc reflection features from the X-ray transient XTE~J1739-285}
\author[Mondal et al.]{\parbox[]{6.5in}{Aditya S. Mondal$^{1}\thanks{E-mail: adityas.mondal@visva-bharati.ac.in}$, B. Raychaudhuri$^{1}$, G. C. Dewangan$^{2}$, Aru Beri$^{3,4}$   \\
\small
$^{1}$Department of physics, Visva-Bharati, Santiniketan, West Bengal-731235, India \\
$^{2}$Inter-University Centre for  Astronomy \& Astrophysics (IUCAA), Pune, 411007 India \\
$^{3}$Indian Institute of Science Education and Research (IISER) Mohali, Punjab 140306, India \\
$^{4}$Physics \& Astronomy, University of Southampton, Southampton, Hampshire SO17 1BJ, UK\\
}}

\date{\today}
\begin{document}
\maketitle
\begin{abstract}
We report on the nearly simultaneous \nicer{} and \nustar{} observations of the known X-ray transient XTE~J1739-285. These observations provide the first sensitive hard X-ray spectrum of this neutron star X-ray transient. The source was observed on 19 February 2020 in the hard spectral state with a luminosity of $0.007$ of the Eddington limit. The broadband $1-70 \kev{}$ \nicer{} and \nustar{} observation clearly detects a cutoff of the hard spectral component around $34-40 \kev{}$ when the continuum is fitted by a soft thermal component and a hard power-law component. This feature has been detected for the first time in this source. Moreover, the spectrum shows evidence for disc reflection -- a relativistically broadened Fe K$\alpha$ line around $5-8 \kev{}$ and a Compton hump in the $10-20 \kev{}$ energy band. The accretion disc reflection features have not been identified before from this source. Through accretion disc reflection modeling, we constrain the radius of the inner disc to be $R_{in}=3.1_{-0.5}^{+1.8}\;R_{ISCO}$ for the first time. In addition, we find a low inclination, $i\sim 33^{0}$. Assuming the magnetosphere is responsible for such truncation of the inner accretion disc above the stellar surface, we establish an upper limit of $6.2\times 10^8$ G on the magnetic field at the poles. 
\end{abstract}
\section{introduction}
Low-mass X-ray binaries (LMXBs) consist of a neutron star~(NS) or a black hole~(BH) accreting from a low-mass ($\leq 1 \Msun$) companion star via Roche-lobe overflow. They may be classified into two categories, the persistent systems and the transient ones, based on their long-term X-ray behavior. Persistent LMXBs are always actively accreting (can show X-ray variability in some cases) whereas transient LMXBs exhibit large swings in their X-ray luminosity. Persistent accretor may have an X-ray luminosity of $L_{X}\geq10^{36} \ergsec{}$ (\citealt{2019ApJ...873...99L, 2017ApJ...836..140L}) whereas, transient systems often undergo cycles of outburst and quiescence due to the modulation in the rate at which matter from the companion star accretes onto the compact object (either NS or BH). Transient LMXBs undergo recurrent bright ($L_{X}\geq 10^{36} \ergsec{}$) outbursts lasting from days to weeks and then return to long interval of X-ray quiescence ($L_{X}\leq 10^{34} \ergsec{}$) lasting from months to years \citep{2010A&A...524A..69D}. The long-term average mass accretion rate of the transient systems are significantly lower than in the persistent systems. \\

XTE~J1739-285 is a transient NS LMXB that was first discovered by the \rxte{} Proportional Counter Array (PCA) in 19 October 1999  \citep{1999IAUC.7300Q...1M}. Since its discovery, the source has shown an irregular pattern of X-ray outbursts. Short outbursts from this source were observed in May 2001 and October 2003 \citep{2007ApJ...657L..97K}. After two years of quiescence, the source became active again in August 2005 \citep{2005ATel..592....1B} and two type-I X-ray bursts were detected with the \inte{}/JEM-X instrument on 30 September 2005 and 4 October 2005 \citep{2005ATel..622....1B}. It confirms that the source harbors an NS as the compact component. The source was detected with \inte{} at a $3-10 \kev{}$ flux of $\sim 2\times 10^{-9} \funit{}$ \citep{2005ATel..592....1B}. However, the flux dropped to $\sim2\times 10^{-10} \funit{}$ nearly a month later \citep{2005ATel..615....1S}. \rxte{} further observed the source on many occasions between 12 October 2005 to 16 November 2005. During this period the flux evolved between $\sim 4\times 10^{-10} \funit{}$ to $\sim 1.5\times 10^{-9} \funit{}$. The source was found to be visible in the \inte{} Galactic Bulge monitoring observations on 9 February 2006 \citep{2006ATel..734....1C}. Since then, outburst activity has been detected in 2012 \citep{2012ATel.4304....1S} and 2019 \citep{2019ATel13148....1B}. Once again, during observations of the Galactic centre region, \inte{} has detected renewed activity of this source on 8 February 2020 \citep{2020ATel13474....1S}. This new outburst cycle was quickly confirmed with a follow-up \swift{}/XRT observation performed on 13 February 2020 \citep{2020ATel13483....1B}. In this outburst cycle \textit{Neutron star Interior Composition ExploreR} (\nicer{}) observed the source on many occasions, and in particular, the observation performed on 13 February 2020 detected 32 X-ray bursts \citep{2021ApJ...907...79B}. \textit{Nuclear Spectroscopic Telescope ARray} (\nustar{}) also observed the source on 19 February 2020, and the spectral analysis of the same is presented in this work. \\ 

From RXTE/PCA observations, \citet{2007ApJ...657L..97K} detected six type-I X-ray bursts and also found the evidence for oscillation at $\sim 1122$ Hz in the brightest X-ray burst. They claimed that this burst oscillation frequency would imply that the source contained the fastest spinning NS known at that time. However, other analyses of the same data found no significant burst oscillation signals (\citealt{2008ApJS..179..360G, 2019ApJS..245...19B}). \citet{2021ApJ...907...79B} did not find any evidence of variability $\sim 1122$ Hz using a sample of \nicer{} data and instead found that the $386.5$ Hz oscillation was the more prominent signal. They concluded that it was unlikely that the source had a submillisecond rotation period. The source exhibits various outburst activities since its discovery, and the extensive timing/burst analysis had been performed in detail by different authors to constrain some crucial parameters of this source. But the spectral analysis has not been performed so far in detail. Therefore, much spectral information could not be extracted from the previous studies. \\
  
In this work, we present the broad-band spectral and variability analysis of nearly simultaneous \nicer{} and \nustar{} observation of this well-known NS X-ray transient it has not been performed so far. We use the high-quality \nicer{} and \nustar{} spectra to study different spectral signatures in detail for the first time. Moreover, we use the joint fit of \nicer{} and \nustar{} data to constrain the hydrogen column density ($N_{H}$) and the reflection composition separately. We focus on studying the accretion geometry for this source by modeling the reflection spectrum. In particular, we aim to constrain the inner disc radius of this LMXB. The paper is structured in the following format: Section 2 presents the observation and data reduction. Section 3 and 4 represent the timing and spectral analysis, respectively. Section 4 provides a discussion of the results obtained from the analyses.\\

\begin{figure*}
\centering
\includegraphics[scale=0.40, angle=0]{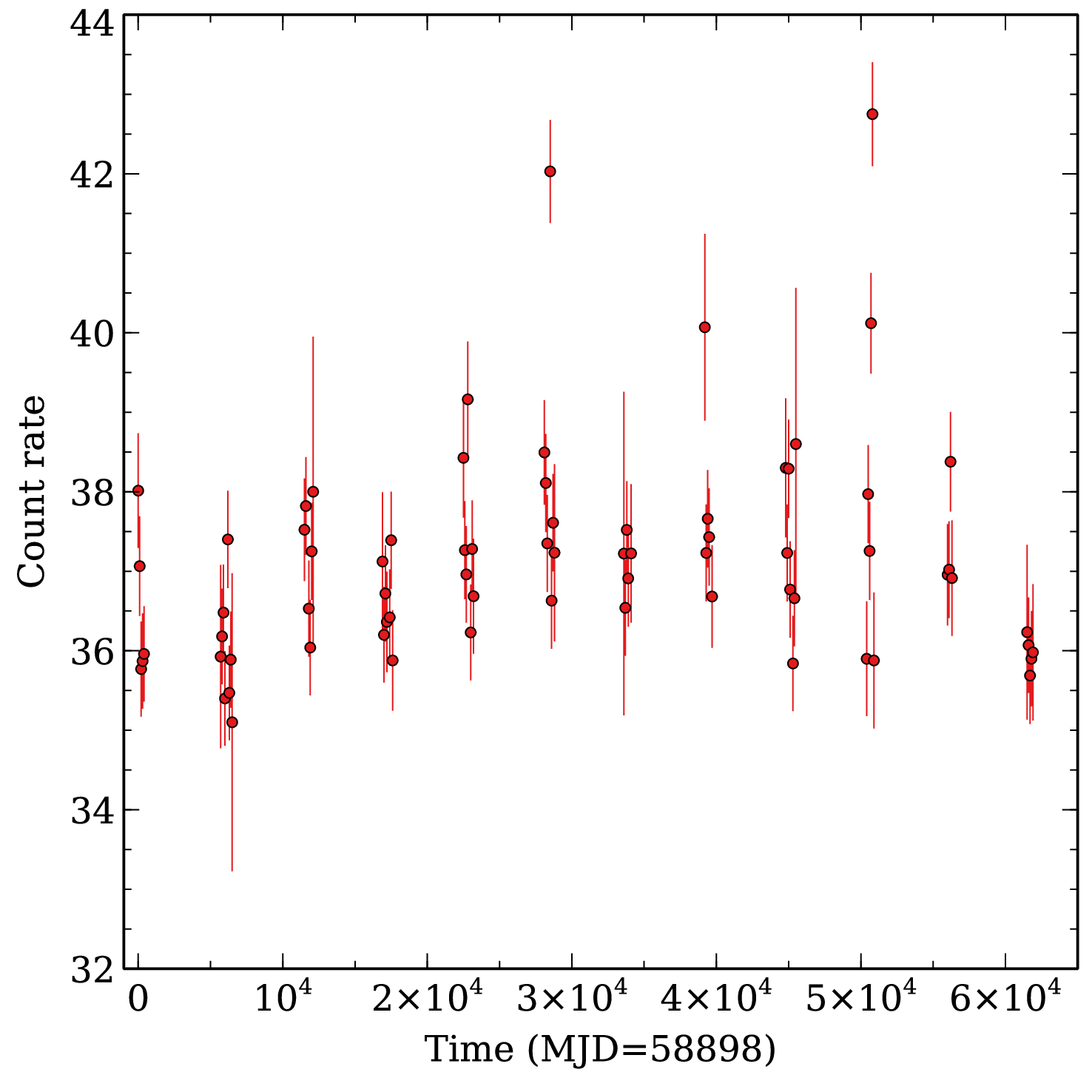}\hspace{2cm}
\includegraphics[scale=0.40, angle=0]{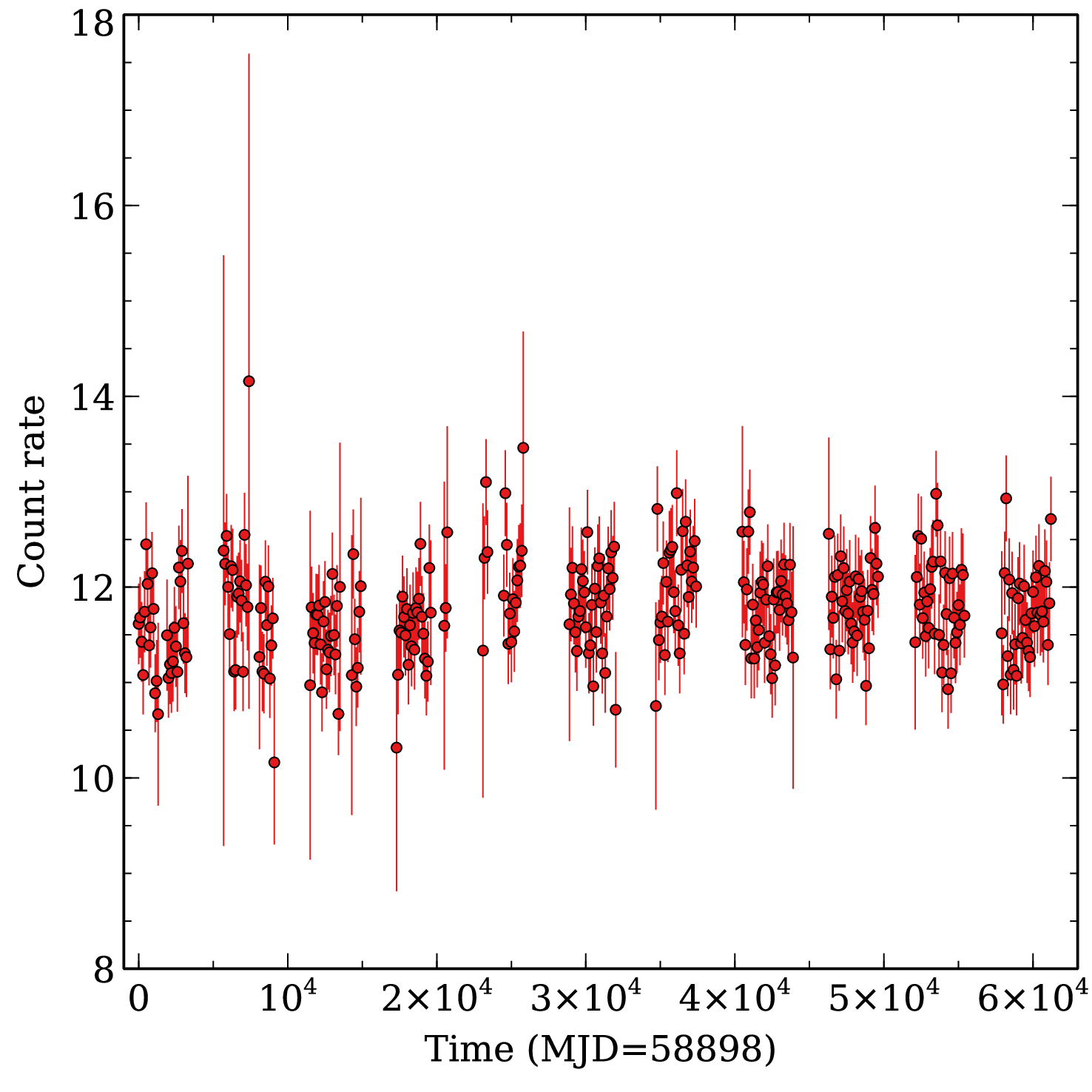}
\caption{Left: $1-10\kev{}$ \nicer{}/XTI light curve of XTE~J1739-285 with a binning of 100 sec. Right: $3-79\kev{}$ \nustar{}/FPMA light curve of the source with 100s binning. Both light curves do not exhibit significant variability in the count rate. } 
\label{Fig1}
\end{figure*}

\section{observation and data reduction}
The X-ray transient XTE~J1739-285 was observed by both satellites the \nicer{} \citep{2016SPIE.9905E..1HG} and the \nustar{} \citep{2013ApJ...770..103H} on February 19, 2020. The \nustar{} observed the source for $29$ ks (Obs ID: 90601307002). Among many observations, \nicer{} has only one observation on the same day with \nustar{}. We have selected this \nicer{} observation (Obs ID: 2050280129) which is nearly simultaneous with the \nustar{} observation. This \nicer{} observation had an exposure time of $6.7$ ks. 

\subsection{\nustar{} data reduction}
The \nustar{} data were collected using two co-aligned grazing incidence hard X-ray imaging focal plane module telescopes FPMA, and FPMB. We processed the \nustar{} data using the \nustar{} data analysis software {\tt NuSTARDAS v2.0.0} for both data sets. During this, we have used the latest calibration file {\tt CALDB v20210524}.  We filtered the event lists using the {\tt nupipeline} tool ({\tt v 0.4.8}). The source events were extracted from a circular region with a radius $120''$ for both  modules FPMA and FPMB centered on the source coordinates. For the background events, we used a circular region of the same radius but far away from the source position for both  instruments. The tool {\tt nuproducts} has been used to create the filtered event files, the background subtracted light curves, the spectra, and the arf and rmf files. During the run of the {\tt nuproducts} tool, we have applied a GTI file which was created by FTOOLS {\tt maketime}. We grouped the FPMA and FPMB spectral data with a minimum of $50$ counts per bin. Finally, The spectra obtained with FPMA and FPMB are fitted simultaneously over the range $3.0-70.0 \kev{}$.

\subsection{\nicer{} data reduction}
We have processed the \nicer{} X-ray timing instrument (XTI) data following the standard steps using the latest {\tt CALDB} {\tt v20210707}. We have reprocessed the data using the {\tt nicerl2} tool, applying standard filtering criteria. Data products (spectra, light curves) have been extracted using the tool {\tt xselect} from the cleaned and screened full array event files. We have used the tool {\tt nibackgen3C50} to extract the background spectrum \citep{2022AJ....163..130R}. The response matrix  $20170601v004$ and ancillary file $20170601v002$ are used from {\tt CALDB}. Finally, the command {\tt grppha} has been used to produce the grouped spectrum with a minimum count of 20 per bin. The \nicer{}/XTI spectrum is fitted over the energy band of $1-10$ \kev{}, considering the impact of low energy noise.

\section{Temporal Analysis}
The light curves for the \nicer{} and the \nustar{} observations are shown in the left and right panel of Figure~\ref{Fig1}, respectively. The $1-10\kev{}$ \nicer{} light curve shows the average count rate of the source $\sim 37$ counts s$^{-1}$. That this source is known for X-ray bursts is corroborated by the \nicer{} light curve showing  the presence of six type-1 X-ray bursts, which have been excluded in this analysis. The $3-79$ \kev{} \nustar{} lightcurve shows an average count rate of  $12$ counts s$^{-1}$. The \nustar{} light curve also detects two type-1 X-ray bursts, and we have excluded those from the present analysis. The light curves suggest that the source mean count rate gradually decreased from $\sim 37$ counts s$^{-1}$ to $12$ counts s$^{-1}$ within a few hours. Further, we run {\tt lcstats} (an XRONOS task) on both light curves to calculate constant source probability through a Kolmogorov-Smirnov (KS) test. We found that, while for the \nustar{} light curve, the average value (taken over the time intervals) of the KS probability of consistency is $\sim 0.168$, it is very small ($<0.05$) for the \nicer{} light curve. It indicates that the \nustar{} light curve does not show significant variability in the count rate, while there can be some variability in the \nicer{} light curve.

\section{spectral analysis}
We have used the spectral analysis package {\tt XSPEC} $v12.11.1$ \citep{1996ASPC..101...17A} to fit the \nicer{} and \nustar{} spectra of this source between 1 to $70$ \kev{}. During the simultaneous fitting of \nicer{}/XTI and \nustar{} FPMA/FPMB data, we have used a model {\tt constant} which coordinates the calibration differences in different instruments in the process of joint fit. We mainly fix the {\tt constant} of \nustar{} FPMA to $1$ and allow the {\tt constant} of \nicer{}/XTI and \nustar{} FPMB to vary. We have modelled the Galactic interstellar medium absorption by the model {\tt TBabs} with {\tt wilm} abundances \citep{2000ApJ...542..914W} and {\tt vern} \citep{1996ApJ...465..487V} cross section. Spectral uncertainties are given at $90$ percent confidence intervals, unless otherwise stated.\\

\begin{figure*}
\centering
\includegraphics[scale=0.40, angle=-90]{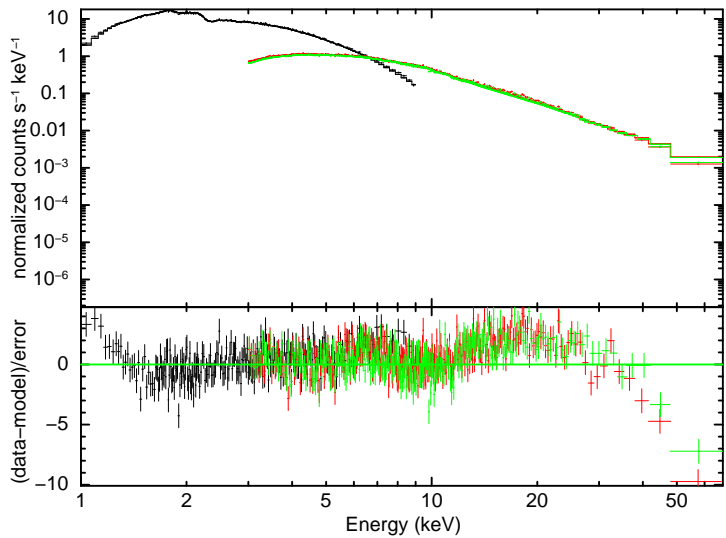}
\caption{Joint fit for the \nicer{} and \nustar{} observations of the source XTE~J1739-285. The residual shows a clear spectral cut-off around $30-40 \kev{}$ when the continuum emission is modelled with an absorbed disc blackbody ({\tt diskbb}) and power-law ({\tt powerlaw)} component. The black points are the \nicer{}/XTI data. The red and green points are the \nustar{} FPMA and FPMB data, respectively. The data were rebinned for plotting purposes.} 
\label{Fig2}
\end{figure*}

\subsection{Continuum modeling}
We initially tried to fit the joint \nicer{} and \nustar{} spectrum by a multicolour disc blackbody component ({\tt diskbb}) to account for the emission from the ion disc and a {\tt powerlaw} component to account for Comptonization i.e. {\tt constant*TBabs*(diskbb+powerlaw)}. This model clearly shows the presence of a hard spectral cut-off $\sim 30-40 \kev{}$ which is evident in Figure~\ref{Fig2}. We, therefore, replaced The {\tt powerlaw} component by a cutoff power-law model {\tt cutoffpl} in {\tt XSPEC}. The continuum is well described by the combination of a {\tt diskbb} and a {\tt cutoffpl} model i.e. {\tt constant*TBabs*(diskbb+cutoffpl)} with $\chi^2/dof=1999.6/1827$. This model yielded a high energy cut-off of $\sim 37 \kev{}$ which may reflect the electron temperature of the corona. Here the power-law component, assumed to be due to the Comptonization process, takes a relatively hard index ($\Gamma=1.16\pm0.04$). For the hard power-law index ($\Gamma<2$), it is well known that the source is in the so-called hard spectral state. We found the absorption column density at $N_{H}=(1.52\pm 0.02)\times 10^{22}$ cm$^{-2}$ which is consistent with \citet{2021ApJ...907...79B}.\\

\begin{figure*}
\includegraphics[scale=0.40, angle=-90]{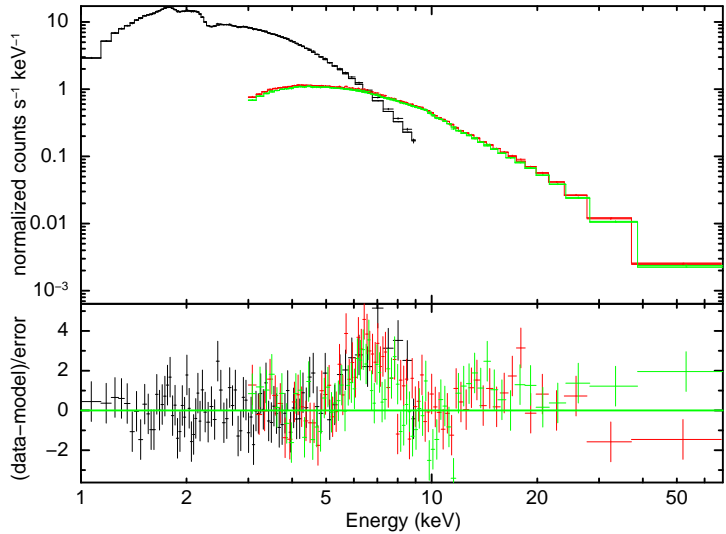}
\caption{The continuum is fitted with the model consisting of a multicolour disk blackbody and a thermal Comptonization model. Model used: {\tt TBabs$\times$(diskbb+nthcomp)}. It revaled un-modelled broad emission line $\sim 5-8$ keV and a hump like feature $\sim 10-20$ keV. The residuals can be indentified as a broad Fe-K emission line and the corresponding Compton hump. The spectral data were rebinned for visual clarity.} 
  \label{Fig3}
   \end{figure*}

To describe the continuum more accurately, we replaced the exponentially cut-off power-law component by the {\tt nthcomp} (\citealt{1996MNRAS.283..193Z, 1999MNRAS.309..561Z}) component as it offers sharper high-energy cut-off and a more accurate low-energy rollover with similar parameters. We explored the possibility that the source of the seed photons in the {\tt nthcomp} component is the disc. This combination {\tt constant*TBabs*(diskbb+nthcomp)} provides a similar description compared to the previous continuum model with $\chi^2/dof=2043.3/1826$. But statistically, an absorbed {\tt cutoffpl} along with a {\tt diskbb} component gives a better description of the continuum emission to this hard state spectrum. The {\tt nthcomp} fit yields the power-law photon index ($\Gamma$), the temperature of the Comptonizing electrons ($kT_{e}$), and the seed photon temperature ($kT_{seed}$) of $\sim 1.72$, $14.6-19.3$\kev{}, and $1.1-3.2$ \kev{}, respectively. The observed high-energy cutoff $\sim 37 \kev{}$ and the $kT{_e}$ in the hard spectrum is consistent with the fact that $E_{cut}\simeq(2-3)\:kT_{e}$. For both continuum model, we found the evidence for fluorescent Fe line emission in the spectrum. Both the continuum model shows similar positive residuals around $5-8 \kev{}$ and $20-30 \kev{}$. These residuals suggest a possible emission line from Fe-K and Compton hump from the reflection of hard X-rays by the cool accretion disc. These features are evident in Figure~\ref{Fig3}.

\subsection{Line modelling: The Fe line region}
\begin{figure*}
   \includegraphics[scale=0.40, angle=-90]{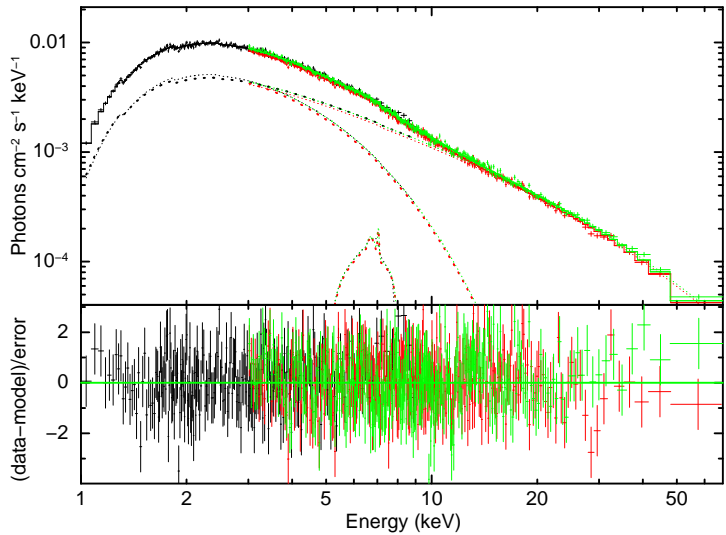}
   
 \caption{The feature $\sim 6.5\kev{}$ is modelled with {\tt relline} model. The overall model used here can be represented as {\tt TBabs$\times$(diskbb+cutoffpl+relline)}. The individual model components are denoted by the dotted lines. Lower panel shows residuals in units of $\sigma$.} 
   \label{Fig4}
   \end{figure*}

To fit the residual observed in the $5-8 \kev{}$ band, we employed the line model {\tt relline} \citep{2010MNRAS.409.1534D}, which assumes an intrinsic zero width emission line transformed by the relevant relativistic effects. The rest energy of the emission is fitted freely within the limit $6.4-6.97 \kev{}$. During this fitting, we have fixed the emissivity index ($\Gamma$) and spin parameter ($a$) to $r^{-3}$ and $0.18$ respectively. The outer radius of the disc is fixed at $1000\;r_{g}$. We adopt $a=0.18$ since the source exhibits a spin frequency of $\sim 386$ Hz \citep{2021ApJ...907...79B} (see below). We set the redshift parameter $z=0$ as it is a Galactic source. The addition of this line model improved the fit significantly with $\chi^2/dof=1852/1823$. The complete model we used here is {\tt constant*TBabs*(diskbb+cutoffpl+relline)}. The corresponding spectra, individual components, and residuals are shown in Figure~\ref{Fig4}. We found the rest energy of the emission line to be $6.57_{-0.10}^{+0.20}$ \kev{}. It predicts that the disc is moderately ionized. Moreover, it predicts a truncated inner accretion disc of radius $\sim 3.7\;R_{ISCO}$, but the disc inclination is found to be pegged at a higher value.

\begin{figure*}
   \includegraphics[scale=0.40, angle=-90]{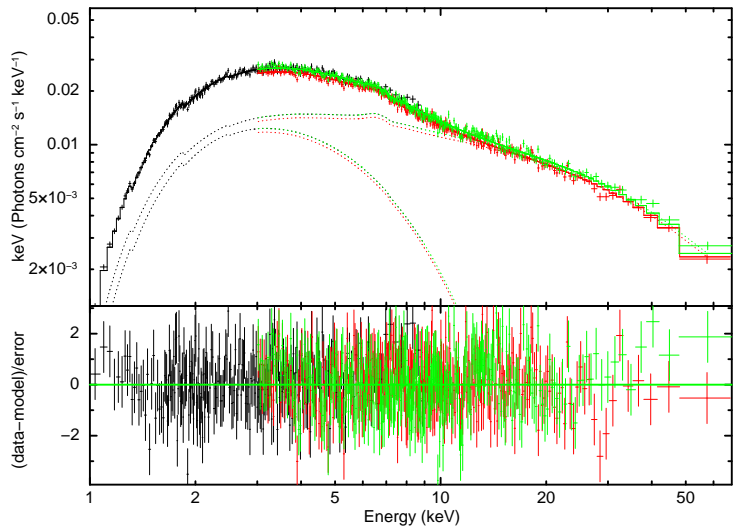}
   
 \caption{Spectrum fitted with the model {\tt diskbb} and {\tt RELXILL}, the best-fit model: {\tt TBabs$\times$(diskbb+RELXILL)}. The individual model components are denoted by the dotted lines. Lower panel shows the ratio of the data to the model in units of $\sigma$. The data were rebinned for plotting purposes.} 
   \label{Fig5}
   \end{figure*}

   \begin{table}
   \begin{center}
\caption{Best-fitting spectral parameters of the 2020 February 19 \nicer{} and \nustar{} observations of the source XTE~J11739-285 using model:  {\tt TBabs$\times$(diskbb+RELXILL)}.}
\vspace{-.5cm}
\begin{tabular}{|p{1.6cm}|p{4.2cm}|p{1.7cm}|}
    \hline
    Component     & Parameter (unit) & Value \\
    \hline
    {\scshape tbabs}    & $N_{H}$($\times 10^{22}\;\text{cm}^{-2}$) &$1.52\pm0.02$     \\
    {\scshape diskbb} & $kT_{in}(\kev)$ &  $1.73\pm 0.04$   \\
    & norm  & $0.74\pm 0.07$     \\
    {\scshape relxill} & $i$ (degrees) & $\leq 33$    \\
    & $R_{in}$($\times R_{ISCO}$) & $3.1_{-0.5}^{+1.8}$\\
    & $log\xi$(erg cm s$^{-1}$) &  $3.54_{-0.30}^{+0.12}$  \\
    & $\Gamma$  & $1.15_{-0.03}^{+0.06} $  \\
    & $A_{Fe}$ ($\times \;\text{solar})$   & $\geq 4.9$  \\
    & $E_{cut}(\kev)$ &  $36.6_{-2.4}^{+3.6}$ \\
    & $f_{refl}$($\times 10^{-2}$)   & $5.59_{-1.10}^{+1.61}$ \\
    & norm ($\times 10^{-4}$)   &  $9.47\pm 0.36$ \\
    & $F^{*}_{total}$ ($\times 10^{-10}$ ergs/s/cm$^2$) & $6.4\pm 0.1$ \\
    & $F_{diskbb}$ ($\times 10^{-10}$ ergs/s/cm$^2$)&  $1.3 \pm 0.1$ \\
    & $F_{relxill}$ ($\times 10^{-10}$ ergs/s/cm$^2$) &  $5.1 \pm 0.2$ \\
    & $L_{1-70 \kev}$ ($\times 10^{36}$ ergs/s) & $1.21 \pm 0.01$ \\	
  
   \hline 
    & $\chi^{2}/dof$ & $1833.12/1822$   \\
    \hline
  \end{tabular}\label{parameters1} \\
  \end{center}
{\bf Note:} The outer radius of the {\tt RELXILL} spectral component was fixed to $1000\;R_{g}$. We fixed emissivity index $q=3$. The spin parameter ($a$) was fixed at $0.2$. \\
$^{*}$All the unabsorbed fluxes are calculated in the energy band $1-70\kev$ using the {\tt cflux} model component. Luminosity is calculated based upon a distance of $4$ kpc \citep{2018AJ....156...58B}.\\

\end{table}

\subsection{self-consistent reflection fitting}
The consideration of reflection off the ion disk is important as the residuals of Figure~\ref{Fig3} show the presence of a broad iron line as well as a reflection hump at high energies. We, therefore, applied the self-consistent relativistic reflection model {\tt RELXILL} \citep{2014ApJ...782...76G} which describes not only the reflection part but also a direct power-law component. The overall model now becomes {\tt constant*TBabs*(diskbb+RELXILL)}. The parameters in the {\tt RELXILL} model are the inner and outer emissivity indices, $q_{1}$ and $q_{2}$, respectively, the break radius, $R_{break}$, between two emissivity indices, the inner and outer radii of the disc, $R_{in}$ and $R_{out}$, respectively, the inclination of the disc, $i$, the spin parameter, $a$, the redshift of the source, $z$, the photon index of the power-law, $\Gamma$, the cut-off energy of the power-law, $E_{cut}$, the ionization parameter, $\xi$, the iron abundance, $A_{Fe}$, the reflection fraction, $r_{refl}$, and the norm which represents the normalization of the model. We have used a single emissivity index $q_{1}=q_{2}=3$. The burst oscillation frequency $\sim 386$ Hz \citep{2021ApJ...907...79B} implies the spin parameter, $a=0.18$ as $a\simeq0.47/P_{ms}$ \citep{2000ApJ...531..447B} where $P_{ms}$ is the spin period in ms. The outer disc radius was fixed at $1000\;R_{g}$. The addition of {\tt RELXILL} model improves the fit significantly to $\chi^2/dof=1833.12/1822$ ($\Delta\chi^2=166$ for the addition of $5$ parameters). The best-fitting parameters of this model are given in Table 1. The corresponding spectra, individual components, and the residuals are shown in Figure~\ref{Fig5}. \\

\begin{figure*}
\includegraphics[scale=0.40, angle=0]{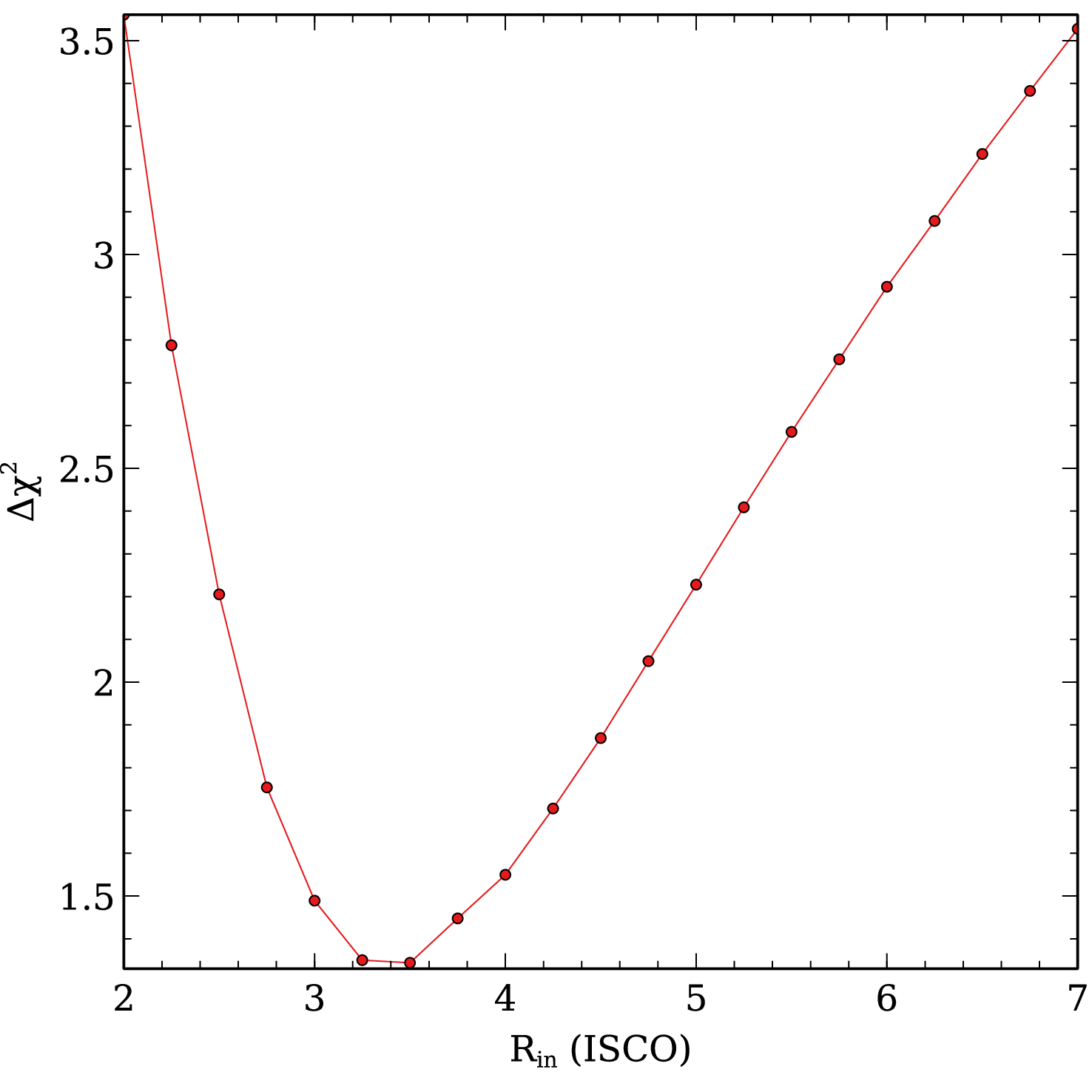}\hspace{2cm}
\includegraphics[scale=0.40, angle=0]{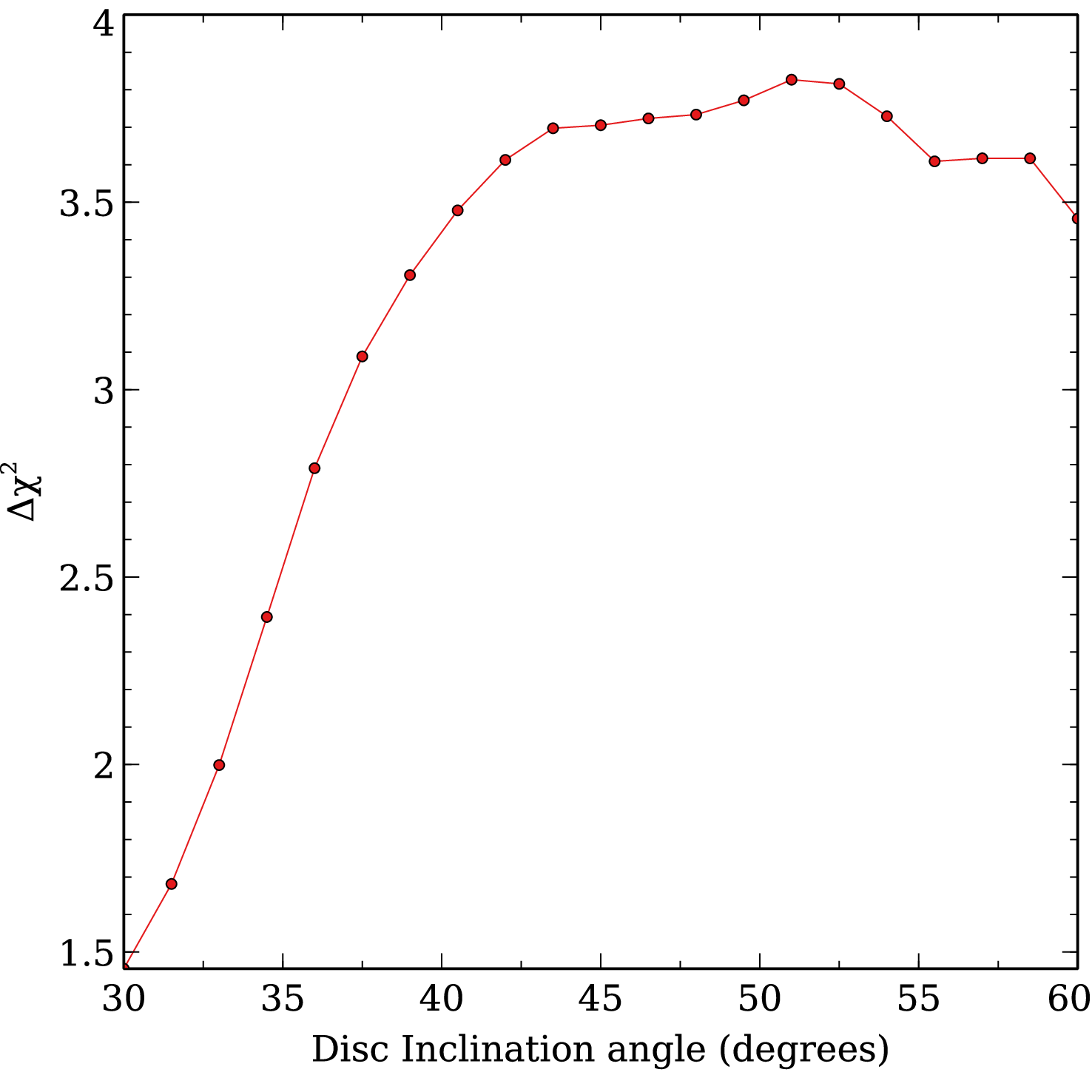}
\caption{Shows change in goodness of fit with respect to the inner disc radius and disc inclination angle. Left panel shows the variation of $\Delta\chi^{2}(=\chi^{2}-\chi_{min}^{2})$ as a function of inner disc radius obtained from the relativistic reflection model ({\tt RELXILL}). Right panel shows the variation of $\Delta\chi^{2}(=\chi^{2}-\chi_{min}^{2})$ as a function of disc inclination angle obtained from the relativistic reflection model. We varied the disc inclination angle between 30 degree and 60 degree.} 
\label{Fig6}
\end{figure*}

We found an inner disc radius to be $3.1_{-0.5}^{+1.8}\;R_{ISCO}$, which implies a significant disc truncation. The inclination is found to be $\leq 33^{0}$. The power-law photon index, $\Gamma$, is $1.15_{-0.03}^{+0.06}$ with a cut-off energy, $E_{cut}$, at $36.6_{-2.4}^{+3.6} \kev{}$. We found a moderate value $3.54_{-0.30}^{+0.12}$ of the disc ionization parameter $\rm{log}\xi$  which is consistent with the typical range observed in different NS LMXBs ($\rm{log}\xi\sim (3-4)$). However, the iron abundance, $A_{Fe}$, is large. The value is greater than $4.9$ times the solar value. We tried to perform the fit after fixing the $A_{Fe}$ at twice the solar value. It did not improve the fit, rather it provided a large error on the position of the inner disc and a high disc inclination angle ($\leq 86^{0}$). However, the other parameters are consistent with the fit that had a free $A_{Fe}$. The overabundance of iron could be indicative of a higher density disc than the hard coded value of $10^{15} \; \rm{cm}^{-3}$ in {\tt RELXILL}. A large iron abundance has already been reported for the LMXB 4U~1636-53 \citep{2017ApJ...836..140L} and 4U~1702-429 \citep{2019ApJ...873...99L} when the reflection component is modeled with {\tt RELXILL}. In all the cases, the source was in the hard spectral state, and the continuum emission was well explained with an absorbed cut-off power-law model. We have used command {\tt steppar} in {\tt XSPEC} to search the best fit for the inner disc radius and inclination for the best-fit model. The left and right panels in Figure~\ref{Fig6} show the $\Delta\chi^2$ of the fit versus the inner disc radius and the disc inclination, respectively, for the best-fit model {\tt constant*TBabs*(diskbb+RELXILL)}. \\

Additionally, we tried to fit the spectrum with {\tt RELXILLCP}, which allows for reflection from a Comptonized disc component. The model {\tt constant*TBabs*(diskbb+RELXILLCP)} did not improve the fit. Moreover, we failed to constrain several important parameters which suggests that the model is not appropriate for this particular observation. Therefore, we do not comment on it further.

\section{Discussion}
We performed a broad-band spectral study for this source with the \nicer{} and \nustar{} observations for the first time. The joint fit of \nicer{} and \nustar{} data allows us to constrain the hydrogen column density ($N_{H}$) and the reflection composition separately. XTE~J1739-285 was in the hard spectral state during these observations. The source was detected with a persistent X-ray flux of $F_{p}=6.4\times 10^{-10}$ erg~s$^{-1}$ cm$^{-2}$ which is consistent with the other observations of $2020$ outburst cycle (\citealt{2021ApJ...907...79B, 2020ATel13483....1B}). The $1-70 \kev{}$ luminosity was $1.21\times 10^{36}$ ergs s$^{-1}$ (assuming a distance of $4$ kpc following \citealt{2018AJ....156...58B}) which is $0.7\%$ of the Eddington luminosity. 
We report, for the first time, different spectral features with \nicer{} and \nustar{} that have not been reported earlier.
 The continuum emission is found to be  well described by a soft thermal component and a hard power-law component with a cut-off $\sim 37 \kev{}$ which has not been detected so far. However, a combined fit of the current \swift{}/XRT and The \inte{} IBIS/ISGRI data (taken on 2020 February 13) showed the evidence of high energy spectral cut-off at $17_{-7}^{+15}\kev{}$ \citep{2020ATel13483....1B}.  Our continuum fit either with the model {\tt cutoffpl} or the {\tt nthcomp} alongwith {\tt diskbb} shows the evidence of Fe emission line $\sim 6.5 \kev{}$ and a Compton hump $\sim 10-20 \kev{}$. Probably these features have been precisely detected for the first time in this source. However, the complex residuals $\sim 6.7 \kev$ have been observed previously by \citet{2007ApJ...657L..97K} using \rxte{}/PCA data and recently by \citet{2020ATel13538....1C} and \citet{2019ATel13148....1B} using the {\it Astrosat}/LAXPC and the \nicer{} data performed on 2020 Ferruary 19-20 and 2019 September 27, respectively. We performed detailed analysis of the resulting reflection spectrum to further probe these features to launch into the following paragraph. \\

The reflection spectrum shows evidence of a significant disc truncation. We measured an inner disc radius of $R_{in}=3.1_{-0.5}^{+1.8} \times R_{ISCO}=(13.1-24.7)R_{g}$ (where $R_{ISCO}=5.05R_{g}$ for a spinning NS) via disc reflection. In addition, the reflection fit yields a low inclination estimate of $\sim 33^{0}$. However, the bursting properties of this source suggest that XTE~J1739-285 is a relatively high inclination system, $65^{0}< i < 90^{0}$, \citep{2021ApJ...907...79B}. The lack of dips or eclipses in the \nicer{} and \nustar{} light curves further indicate that we are not viewing the system edge-on, allowing an upper limit on the inclination of $i\le 75^{0}$ \citep{1987A&A...178..137F}. We note that the inclination obtained with reflection spectroscopy is for the inner accretion disc and can possibly be different from the binary inclination \citep{2020ApJ...899...44W}. \\

In the present work, we have shown  that the disc of XTE~J1739-285 is truncated well above the stellar surface. A truncated disc has not been indicated for this source in the prior investigations. The sensitive and pile-up free \nustar{} spectrum of this source allows us to put such a strong radius constraint. The inferred inner disc radius seems consistent with the observed low luminosity (low/hard state). This kind of larger inner disc radius of $\sim 15-30\; R_{g}$ has been observed before for a number of other moderate-luminosity, intermittent NS LMXBs (\citealt{2016A&A...596A..21I, 2016ApJ...819L..29K, 2013MNRAS.429.3411P, 2011ApJ...731L...7M, 2016MNRAS.457.2988P}). Below we discuss some of the possibilities leading to the disc truncation.\\

The disc truncation is likely the result of either a state transition associated with a receding disk, a boundary layer, or a magnetic field exerting pressure on the disc. \citet{1997ApJ...489..865E} suggests that a receding disc from a state transition is typically associated with low-luminosity and a hard power-law dominated X-ray spectra. The XTE~J1739-285 spectra presented here are observed in the hard, low luminosity state ($\sim 1.21\times 10^{36}$ ergs s$^{-1}$). The continuum emission at energies of $>8 \kev{}$ is strongly dominated by the hard power-law component. Thus, a state transition may play a significant role in the disc truncation as the source shows irregular pattern of X-ray outbursts, and its flux evolved considerably over time. To establish the state transitions in the system, multiple observations of the outburst cycle akin to the one observed in 2020 are required.\\

We examined different scenarios responsible for disc truncation. We first tested whether the boundary layer is responsible for the disc truncation by calculating its maximum radial extension. To do so, we begin with an estimation of the mass accretion rate per unit area, using Equation (2) of \citet{2008ApJS..179..360G}
\begin{equation}
\begin{split}
\dot{m}=&\:6.7\times 10^{3}\left(\frac{F_{p}\:c_\text{bol}}{10^{-9} \text{erg}\: \text{cm}^{-2}\: \text{s}^{-1}}\right) \left(\frac{d}{10 \:\text{kpc}}\right)^{2} \left(\frac{M_\text{NS}}{1.4 M_{\odot}}\right)^{-1}\\
 &\times\left(\frac{1+z}{1.31}\right) \left(\frac{R_\text{NS}}{10\:\text{km}}\right)^{-1} \text{g}\: \text{cm}^{-2}\: \text{s}^{-1}.
 \end{split} 
\end{equation}
This yields a mass accretion rate of $2.2\times 10^{-10}\;M_{\odot}\;\text{y}^{-1}$ at a persistent flux $F_{p}=6.4\times 10^{-10}$ erg~s$^{-1}$ cm$^{-2}$, assuming the bolometric correction $c_{bol} \sim 1.38$ for the nonpulsing sources \citep{2008ApJS..179..360G}. In this equation we assume $1+z=1.31$ (where $z$ is the surface redshift) for a NS with mass ($M_{NS}$) 1.4 $M_{\odot}$ and radius ($R_{NS}$) $10$ km. The estimated mass accretion rate is consistent with \citet{2021ApJ...907...79B}. At this mass accretion rate, using Equation (2) of \citet{2001ApJ...547..355P}, we estimated the maximum value of the boundary layer to extend to $R_{B}\sim 5.3\;R_{g}$ (assuming $M_{NS}=1.4\:M_{\odot}$ and $R_{NS}=10$ km). The actual value may be larger than this if we consider the changes in viscosity and rotation of this layer. Still, the radial extension of the boundary layer is somewhat smaller than the disc truncation radius. Thus, it is implausible that the boundary layer is responsible for the disc truncation. \\

Secondly, we entertain the possibility that the magnetic field associated with the NS would be responsible for the dsc truncation \citep{1975A&A....39..185I}. An upper limit of the magnetic field strength of the NS can be estimated with the inferred inner disc radius. We used Equation (1) of \citet{2009ApJ...694L..21C} to calculate the magnetic dipole moment ($\mu$)
\begin{equation}
\begin{split}
\mu=&3.5\times 10^{23}k_{A}^{-7/4} x^{7/4} \left(\frac{M}{1.4 M_{\odot}}\right)^{2}\\
 &\times\left(\frac{f_{ang}}{\eta}\frac{F_{bol}}{10^{-9} \text{erg}\: \text{cm}^{-2}\: \text{s}^{-1}}\right)^{1/2}
 \frac{D}{3.5\: \text{kpc}} \text{G}\; \text{cm}^{3},
\end{split} 
\end{equation}
where $\eta$ is the accretion efficiency in the Schwarzschild metric, $f_{ang}$ is the anisotropy correction factor. The geometrical coefficient $k_{A}$ depends on the conversion from spherical to disc accretion (numerical simulation suggests $k_{A}=0.5$ whereas the theoretical model predicts $k_{A}<1.1$). We note that \citet{2009ApJ...694L..21C} modified $R_{in}$ as $R_{in}=x\:GM/c^{2}$. We calculated a bolometric flux $F_{bol}\approx 7.45\times 10^{-10}$ \funit{} by extrapolating the best-fit over the $0.1-100$ \kev{} range. Utilizing the upper limit of the measured inner radius ($R_{in}\leq 24.7\;R_{g}$) from the reflection model, a mass of $1.4\:M_{\odot}$, a radius of $10$ km, and a distance of $4$ kpc, we found an upper limit of the magnetic field strength of $B\leq 6.2\times10^{8}$ G at the magnetic poles, assuming $k_{A}=1$, $f_{ang}=1$ and $\eta=0.1$. It may be noted that we have retained similar assumptions regarding the geometrical and efficiency parameters as \citet{2009ApJ...694L..21C}. \\

To further investigate the disc truncation scenario, we calculate the position of the magnetospheric radius ($R_{M}$). During the process of accretion, the magnetic field truncates the geometrically thick accretion disc near this point. The accretion disc is interrupted at the magnetospheric radius for disc-accretion, given by \citep{1979ApJ...232..259G}
\begin{equation}
R_{M}=1300\:L_{37}^{-2/7}\:M^{1/7}\:R_{6}^{10/7}\:B_{12}^{4/7} \;\text{km},
\end{equation}
where $M$ is the mass of NS in $1.4\:M_{\odot}$ units, $R_{6}$ is the radius in units of $10^{6}$ cm, $B_{12}$ is the surface magnetic field strength in $10^{12}$ G units, and $L_{37}$ is the accretion luminosity in units of $10^{37}$ erg~s$^{-1}$. If we take a $B$ field upper limit of $\sim 6.2\times 10^{8}$ G (estimated in the previus section) and adopt $1.2\times 10^{36}$ erg~s$^{-1}$ as the luminosity of the source, the implied magnetospheric radius is $\sim 45$ km from the central object (assuming $M_{NS}=1.4\:M_{\odot}$ and $R_{NS}=10$ km). This is in good agreement with the position of $R_{in} \:(27-52 \:\text{km})$. \\

We have discussed some of the possibilities of disc truncation mechanism. We find that the radial extent of the boundary layer is smaller than the disc truncation radius. So, this possibility of disc truncation can be ruled out. Moreover, this single observation cannot confirm whether a state transition is responsible for disc truncation. But trancation by the magnetosphere can not be ruled out as the position of the magnetospheric radius is consistent with the position of the inner disc. However, multiple observations of different mass accretion rates may be helpful in determining the definitive truncation mechanism for this system.

\section{Acknowledgements}
We thank the annonymous referee for the comments, which have improved this work much. This research has made use of data and/or software provided by the High Energy Astrophysics Science Archive Research Centre (HEASARC). This research also has made use of the \nustar{} data analysis software ({\tt NuSTARDAS}) jointly developed by the ASI science center (ASDC, Italy) and the California Institute of Technology (Caltech, USA). This work is supported by NASA through the NICER mission. ASM and BR would like to thank Inter-University Centre for Astronomy and Astrophysics (IUCAA) for their facilities extended to them under their Visiting Associate Programme. 

\section{Data availability}
Both observational data sets with Obs. IDs $90601307002$ (\nustar{}) and $2050280129$ (\nicer{}) dated February 19, 2020 are in public domain put by NASA at their website https://heasarc.gsfc.nasa.gov.

\def\apj{ApJ}
\def\apjl{ApJl}
\def\pasp{PASP} \def\mnras{MNRAS} \def\aap{A\&A} \def\physerp{PhR} \def\apjs{ApJS} \def\pasa{PASA}
\def\pasj{PASJ} \def\nat{Nature} \def\memsai{MmSAI} \def\araa{ARAA} \def\iaucirc{IAUC} \def\aj{AJ} \def\aaps{A\&AS}
\bibliographystyle{mn2e}
\bibliography{aditya}

\end{document}